\documentstyle[11pt]{article}

\newcommand{\be}{\begin{equation}}
\newcommand{\ee}{\end{equation}}
\newcommand{\bea}{\begin{array}}
\newcommand{\ea}{\end{array}}
\newcommand{\beqa}{\begin{eqnarray}}
\newcommand{\eeqa}{\end{eqnarray}}
\newcommand{\bean}{\begin{eqnarray*}}
\newcommand{\eean}{\end{eqnarray*}}

\def\BI{{\rm 1\!l}}
\def\vonex{\matrix{{}\cr v(x) \cr {}^1}}

\def\vtwoy{\matrix{{}\cr v(y) \cr {}^2}}

\def\gone{\matrix{{}\cr g \cr {}^1}}
\def\gtwo{\matrix{{}\cr g \cr {}^2}}

\def\up#1{\leavevmode \raise.16ex\hbox{#1}}

\newcommand{\journal}[4]{{\sl #1 }{\bf #2} \up(19#3\up) #4}

\newcommand{\gapproxeq}{\lower
 .7ex\hbox{$\;\stackrel{\textstyle >}{\sim}\;$}}
\newcommand{\lapproxeq}{\lower .7ex\hbox{$\;\stackrel
{\textstyle <}{\sim}\;$}}

\newcounter{appendice}

\def\thebibliography#1{{\bf REFERENCES\markboth
 {REFERENCES}{REFERENCES}}\list
 {[\arabic{enumi}]}{\settowidth\labelwidth{[#1]}\leftmargin\labelwidth
 \advance\leftmargin\labelsep
 \usecounter{enumi}}
 \def\newblock{\hskip .11em plus .33em minus -.07em}
 \sloppy
 \sfcode`\.=1000\relax}

\begin{document}

\begin{flushright}

March 1999\\
\end{flushright}
%\vspace*{5mm}

\centerline{ \LARGE   T-Duality for Coset Models}
\vskip 2cm

\centerline{ {\sc  A. Stern }  }

\vskip 1cm

\centerline{  Dept. of Physics and Astronomy, Univ. of Alabama,
Tuscaloosa, Al 35487, U.S.A.}

\vskip 2cm

\vspace*{5mm}

\normalsize
\centerline{\bf ABSTRACT}

We        construct  dual Lagrangians
for    $G/H$ models in two space-time dimensions
for arbitrary Lie groups $G$ and $H\subset G$.
Our approach does not require choosing coordinates on $G/H$, and allows
for a natural generalization to Lie-Poisson duality.
For the case where the
target metric on $G/H$ is induced from the invariant  metric on $G$,
the dual system is a gauged Higgs model, with a nonconstant metric and
a coupling to an antisymmetric tensor.  The  dynamics for the
 gauge connection is governed by a $BF$-term.   Lie-Poisson duality is
relevant once we allow for a more general class of target metrics,
as well as for couplings to an antisymmetric tensor,
 in the primary theory.  Then the dual  theory is
 written on a  group $\tilde G$ dual to $G$,
and the gauge group $H$  (which, in general, is not a subgroup of
 $\tilde G$)
acts nonlinearly on $\tilde G$.  The dual system  therefore gives a
 nonlinear realization of a gauge theory.  All dual descriptions
are shown to be canonically equivalent to the corresponding primary
descriptions, at least at     the level of the current algebra.

\vskip 2cm
\vspace*{5mm}

\newpage
\scrollmode

\section{Introduction}

Cosets models have been around for a long time.  There
 are many well known examples, including
 the $O(3)$ nonlinear $\sigma$-model,
 the $SU(N)\times
SU(N)$ chiral model and $CP^{n-1}$ model.
The standard Lagrangian $L$ for a
  $G/H$ model contains   terms quadratic in velocities, and
  can be  written on the tangent manifold
 $TG$ of $G$.  It is invariant under local transformations
by $H\subset G$, and also possibly global transformations by $G$.
Our interest here concerns the question of possible dual
Lagrangians.
Motivated by developments in string theory, techniques have been
 devised for finding dual Lagrangian descriptions of field theories
 in two space-time dimensions.  Here we  refer
more specifically to the techniques of nonabelian T-duality, as they
 can be applied  to $G/H$ models.\cite{rtd}  In the original approach,
 one makes a choice of coordinates on the manifold, whereby isometry
directions are distinguished from non-isometry directions,
  and then  introduces a gauge connection for every  isometry.
   This then means that  various coset manifolds must be analyzed
 separately, and more often than not,
 the analysis may be quite
formidable.  It  therefore seems safe to say
  that following such an approach,
  the general structure of the dual theory
for  $G/H$ models is not very transparent.

In this article, we instead write down a dual Lagrangian $\tilde L$ for
$G/H$ models  without relying on coordinate charts
  on $G/H$,  and we check that it leads to the same dynamics as the
standard Lagrangian.  The result applies for arbitrary Lie groups $G$ and
 $H$, and furthermore, we are
 able to do this for a variety of  target space  metrics.
We show that at the level of currents, the  dual
descriptions are   canonically equivalent to the standard ones.
 Our approach differs
from previous ones\cite{dwzw} which realize coset models as
gauged Wess-Zumino-Witten models\cite{gwzw}.  It is more closely
related to the treatments in refs. \cite{klm96}, \cite{sf2} and
\cite{prwo},  which are based on Lie-Poisson T-duality.\cite{ks95}

We now summarize some of the features of the dual Lagrangian
 $\tilde L$.
While the standard Lagrangian $L$, which we also refer to as the primary
Lagrangian,  can   be written without the use of gauge connections or
topological terms,  connections, as well as topological terms,
 enter    in the  dual description.  In $\tilde L$ the dynamics
of gauge fields   is governed by a    $BF$  term.
To illustrate some of the structures of the dual theory, we first consider
 the  example of  $G=SU(2)$ and $H=U(1)$, where
one recovers  the familiar dynamics of the $O(3)$ nonlinear
$\sigma$-model.  Its dual theory
can be written in terms of a scalar  doublet  coupled to a $U(1)$
gauge field, whose field strength is proportional to the instanton
 number.
The scalars fields are also coupled to a nonconstant metric and an
 antisymmetric tensor.  It is convenient to also introduce
an auxiliary scalar field in the dual description,
which then allows for a local  expression
for the metric.  The auxiliary  field couples to the $U(1)$ connection,
as  well.    By choosing a gauge and eliminating the auxiliary variables
it should be possible to recover all previous expressions for the dual
Lagrangian.
     If in $\tilde L$ one introduces the a kinetic energy term
for the gauge field,  this corresponds
to adding a higher order term in the primary Lagrangian $L$.
There is some recent interest in such theories in connection with
knot solitons.\cite{knot}

For the case of arbitrary $G$ and $H$, with the constant group metric
appearing  in   the primary Lagrangian
 $L$, the dual system  is a gauged Higgs theory   with a
       nonconstant    metric and coupling to an antisymmetric tensor.
 Now there are  $dim[G/H]$ scalar fields, as well as
 $dim[H]$ auxiliary fields, coupled to  $dim[H]$  gauge connections.
Once we drop the restriction  of  constant target metrics
 in the primary theory, as well as allow for a
coupling an antisymmetric tensor, a more
interesting structure emerges in the dual theory.  The latter
is now  written on a  group $\tilde G$ dual to $G$
(in the Lie-Poisson sense),
and the gauge group $H$  (which, in general, is not a subgroup of
 $\tilde G$)  has a  nonlinear  action on $\tilde G$.
As a result, one ends up expressing the dual Lagrangian $\tilde L$
in terms of nonlinear covariant derivatives on
$\tilde G$, and the  gauge symmetry is not immediately obvious.
But our construction of $\tilde L$, starting
from the Hamiltonian formalism, ensures gauge invariance,
and thus provides a
technique for constructing nonlinear realizations of gauge theories.

In section 2,  we  review the standard Lagrangian
              description of $G/H$ models, and then specialize to the
$O(3)$ nonlinear $\sigma$-model in section 3.  There, we construct the
dual theory, and demonstrate that the role of identities and equations
of motion are interchanged in the two formulations.  The analysis
is generalized to the case of arbitrary $G$ and $H$ in section 4.
In section 5, we construct the corresponding Hamiltonian descriptions
 of the two
 theories and demonstrate that, at least locally, the two theories are
canonically equivalent.   We also show that these theories are
canonically equivalent to an equivalence class of
Kac-Moody algebras
associated with the cotangent bundle group $T^*G$.
 In section 6, following \cite{prwo},\cite{Sfet},
we generalize to  Kac-Moody algebras  associated with the  Drinfeld
double group \cite{drin}.  This leads to Lie-Poisson T-duality,
\cite{ks95} and allows for nonconstant
metrics and antisymmetric tensors in both the primary and dual
 descriptions.

\section{Standard Lagrangian Description}
\setcounter{equation}{0}

We now introduce our notation, and
 review the standard Lagrangian
description of $G/H$ models.

 Say $G$ and $H\subset G$ are $N$ and $N-M$ dimensional groups,
respectively, with
the former generated by $e_i, \;i=1,2,..N$, and
 having commutation relations:
\be  [e_i,e_j]=c_{ij}^k e_k          \;. \label{crG}      \ee
We can split the generators into  $e_a,$ $a=1,2,...M$ and
  $\hat{e}_\alpha=e_{M+\alpha},$ $\alpha=1,2,...N-M$,
   the latter generating $H$,
\be     [\hat{e}_\alpha,\hat{e}_\beta]=\hat{c}_{\alpha \beta}^\gamma
\hat{e}_\gamma          \;  ,
 \qquad   \hat{c}_{\alpha\beta}^\gamma =c_{M+\alpha
  \;\;M+\beta}^{M+\gamma  }    \;.\ee
 We will  assume
that the metric ${\tt g}_{ij}$ on $G$ is nondegenerate and block
diagonal, i.e.
\be     {\tt g}_{a\; M+\alpha}  =0 \;.\label{metr}\ee
The structure constants satisfy
\be c_{M +\alpha\;M +\beta}^c =0\;,\qquad
c_{M +\alpha\;b}^{M +\gamma}
 =0 \;.\label{stco}\ee   The
second relation in (\ref{stco}) follows
 from the first, using (\ref{metr}) and
 the invariance property
$c^i_{jk}{\tt g}_{i\ell} ={\tt g}_{ji}c^i_{k\ell}$.

We can now  give the  expression for the primary Lagrangian density $L$.
We take the fundamental fields   $g(x)$  to  have values
 in $G$.  $L$ is a quadratic function  of the velocity
 components  $( g^{-1}\partial_\mu g)^a$,
where  $\mu=0,1$ is the space time index and
$ g^{-1}\partial_\mu g=
    ( g^{-1}\partial_\mu g)^ie_i $.
Utilizing the group metric  ${\tt g}_{ij}$ projected onto
 $G/H$, and following\cite{bgm},\cite{book},  we write
\be L=        -
\frac\kappa {2} {\tt g}_{ab} ( g^{-1}\partial_\mu g)^a
 (g^{-1}\partial^\mu g)^b
 \;,  \label{gstLagr}\ee
 where $a,b=1,2,...M$ and    $\kappa$ is an arbitrary constant.
$L$  is  gauge invariant under
 \be  g(x)\rightarrow g(x) h(x)\;, \quad h(x)\in H\;,\label{ggt}\ee
and consequently defines a theory on $G/H$.  $L$ is also
 invariant under global transformations
 \be g\rightarrow g_0g       \;,\qquad g_0 \in G  \;.\label{glblsym}\ee
For the case $G=SU(2)$ and $H=U(1)$ we obtain the dynamics of
the $O(3)$ nonlinear $\sigma$-model.  $G=SU(n)\times SU(n)$ and
 $H=SU(n)_{diagonal}$ yields the  $G=SU(n)\times SU(n)$  chiral model,
while    $G=U(n)$ and $H=U(n-1)\times U(1)$  corresponds to the
 $CP^{n-1}$ model.  In  sections 3-5, we shall discuss the
 dual description of all such
models in two space-time dimensions.

In section 6,
we obtain the dual for the more general system obtained  by replacing
 ${\tt g}_{ab}$ in (\ref{gstLagr})
   by a nonconstant metric $\gamma(g)_{ab}$.
 We further consider the addition
of a term  with a coupling to an antisymmetric tensor $\rho(g)_{ab}$.
 The primary Lagrangian then has the form
\be L=        -
\frac\kappa {2} \gamma(g)_{ab} ( g^{-1}\partial_\mu g)^a
 (g^{-1}\partial^\mu g)^b   -
\frac\kappa {2} \epsilon^{\mu\nu} \rho(g)_{ab} ( g^{-1}\partial_\mu g)^a
 (g^{-1}\partial_\nu g)^b      \;.  \label{plLagr}
 \ee
Now gauge invariance under (\ref{ggt}) follows  only
for certain metrics  $\gamma(g)_{ab}$ and antisymmetric tensors
$\rho(g)_{ab}$, i.e. those which gauge transform under the adjoint action
of $H$,
\be \gamma(gh)_{ab} = a(h)_a^{\;\;c} \;  a(h)_b^{\;\;d}\; \gamma(g)_{cd}
 \;,\label{togam}\ee
where   $ \;   he_a h^{-1}= a(g)_a^{\;\;b}e_b \;,$ and we have the same
transformation property for $\rho(g)$.  Thus the target space is
again $G/H$.
On the other hand, the global symmetry (\ref{glblsym}) is broken.
Starting with the Hamiltonian formalism, we shall
 obtain a class of
metrics   and antisymmetric tensors consistent with the gauge symmetry,
and for every such metric   and antisymmetric tensor  there is
 a corresponding dual Lagrangian description.

\section{ $O(3)$ nonlinear $\sigma$-model}
\setcounter{equation}{0}

\subsection{Primary Formulation}

The simplest nontrivial example of the above system is
the $O(3)$ nonlinear $\sigma$-model.  The   target  space in this case is
   $S^2$.     Thus we    can introduce
the  fields $\psi^i(x)$, $i=1,2,3$,
satisfying the constraint $\psi^i(x)\psi^i(x) =1$.
  The standard Lagrangian density for the $O(3)$ nonlinear $\sigma$-model
  is
\be L=-\frac\kappa{2} \partial_\mu \psi^i
   \partial^\mu\psi^i \;.\label{orLa}
\ee
Alternatively, we can  rewrite the Lagrangian according to
 (\ref{gstLagr}),
with $G=SU(2)$ and $H=U(1)$.    For this we first realize
  the constraint  by writing
\be      \psi^i(x)e_i= g(x)e_3 g(x)^{-1} \;,\label{psitg}\ee  $g(x)$
being $SU(2)$-valued fields and
 $e_i$ being $SU(2)$ generators.
Since $\psi^i$ are  invariant under
 \be g(x) \rightarrow g(x) e^{\lambda(x) e_3}\;,\label{gt}\ee
 we have thus  introduced  a  $U(1)$ gauge symmetry, the generator being
  $\hat{e}_1=e_3$.
The structure constants and invariant metric  for $SU(2))$ are
 $c_{ij}^k=\epsilon_{ijk}$ and  ${\tt g}_{ij}=\delta_{ij}$.
It is not hard to show\cite{bgm},\cite{book} that the Lagrangian density
  (\ref{orLa})  may be reexpressed according to
  \be L= -
\frac\kappa{2} ( g^{-1}\partial_\mu g)^a   (g^{-1}\partial^\mu g)^a \;,
\label{stLagr}\ee  where $    a=1,2    $.
The Lagrangian, like $\psi^i$, is  invariant under
(\ref{gt}), and the
physical degrees of freedom therefore span $ SU(2)/U(1) \simeq   S^2 $.

The equations of motion resulting from variations of $g$ imply the
existence of  conserved currents $ \partial_\mu J^{i\mu}=0$, where
\be  J^{i\mu} = [a(g)_a^{\;\;i}  (g^{-1}\partial^\mu g)^a] \;,
\label{curcon}\ee
where   $ a(g)$ denotes  the adjoint matrix representation of   $g$,
$ \;   ge_i g^{-1}= a(g)_i^{\;\;j}e_j \;.$   Due to the gauge symmetry
there are only two equations in (\ref{curcon}).  To see this, let us
rewrite the equations in terms of a new set of variables
$\pi_\mu ^ a$ and  $ A_\mu$:
  \beqa \pi_\mu ^ a & =&\epsilon_{\mu\nu}
 ( g^{-1}\partial^\nu g)^a \;,\label{defpi} \\
  A_\mu &=&(g^{-1}\partial_\mu g)^3\;.
\label{defA}      \eeqa
$A_\mu$  transforms as a $U(1)$ connection under (\ref{gt}), while
$\pi_\mu ^ a $ undergo a rotation in the internal space.  We can
  define   the covariant derivative for the latter variables according to
          \be (D_\mu\pi_\nu)^a=\partial_\mu\pi_\nu ^a -
          \epsilon_{ab} A_\mu \pi_\nu ^ b\;,\ee
           where    $   \epsilon_{ab}  =\epsilon_{ab3}\;. $
Now by using $
\partial_\mu  a(g)_i^{\;\;j}= \epsilon_{ik\ell} \; a(g)_k^{\;\;j}
( g^{-1}\partial_\mu g)^\ell \;,$ we see
 that the equations of motion state that the covariant curl of $\pi^a$ is
  zero:
\be \epsilon^{\mu\nu}(D_\mu \pi_\nu)^a =0   \;.
\label{o3eom}\ee
In addition to the equations of motion (\ref{o3eom}),
 we have a set of  three Maurer-Cartan equations.  In terms of
 $A_\mu$ and $ \pi_\mu ^a$, these three identities can  be expressed as
\beqa
 (D^\mu\pi_\mu)^a & =&0\;,\label{codiv} \\
F& =&{1\over 2}\epsilon_{ab}  \epsilon^{\mu\nu}
\pi_\mu^a\pi_\nu^b\;,\label{do3eom} \eeqa
$F$ being the $U(1)$ curvature,
$ F=\epsilon^{\mu\nu} \partial_\mu A_\nu \;.  $  When expressed in terms
of the fields on $S^2$, it is given by
  \be F= 4\pi\;\biggl( -\frac 1{8\pi}  \epsilon_{ijk}
   \epsilon^{\mu\nu} \psi^i\partial_\mu \psi^j\partial_\nu \psi^k
     \biggr)\;,\label{instfld}\ee
the quantity in parenthesis being the instanton number density.

 In summary, the dynamical system is determined by
 the equations of motion (\ref{o3eom}), as well as
  identities
 (\ref{codiv}) and (\ref{do3eom}).   We note that
 there are not an equal number of equations of motion and
identities.   We next
present a  dual formulation of the above system, where
 eqs. (\ref{codiv}) and (\ref{do3eom})
 are equations of motion and eq. (\ref{o3eom}) follows from
  an identity
(evaluated on-shell).

\subsection{Dual Formulation}

The dual Lagrangian $\tilde L$
 will be expressed in terms of three unconstrained
  scalar fields  $\theta$ and
$\chi_a$, $a=1,2$, and a $U(1)$ connection $A_\mu$.  $\theta$ ends
up playing the role of an auxiliary variable.
Before considering the full system,
we first examine a Lagrangian $\tilde L_0$ depending only on
 $\chi_a$ and  $A_\mu$:
\be \tilde L_0=
-{\alpha\over 2} (D_\mu \chi)_a  (D^\mu \chi)_a- {\beta\over 2}
\epsilon^{\mu\nu}\epsilon_{ab} (D_\mu \chi)_a (D_\nu \chi)_b
\;,\ee
where   the covariant derivative $(D_\mu \chi)_a $
is defined by
$    (D_\mu \chi)_a      = \partial_\mu \chi_a - \epsilon_{ab} \chi_b
A_\mu\;. $
$\tilde L_0$  is gauge invariant under the following
infinitesimal variations: \beqa
 \delta \chi_a & =&\lambda
 \epsilon_{ab} \chi_b \cr     \delta A_\mu & =&\partial_\mu
 \lambda \;,\label{dgt}\eeqa
   where here we take $\lambda$ to be infinitesimal.
   $\alpha\delta_{ab}$ and $\beta\epsilon_{ab}$ represent the dual
    metric and antisymmetric
    tensor, respectively.
 Upon assuming for the moment that $\alpha$ and $\beta$ are constants,
  we get
 (\ref{codiv}) as the equations of motion associated  with variations of
   $\chi_a$,
 provided we now define  $\pi_\mu^a$  (up to an overall constant)  by
\be  \pi_\mu^a   =
-\alpha (D_\mu \chi)_a-\beta
\epsilon_{\mu\nu}\epsilon_{ab}  (D^\nu \chi)_b \;. \label{ddefpi}
\ee
 This definition
leads to the identity \be \epsilon_{ab}  (D^\mu \chi)_a \pi_\mu^b =0
\;.\label{did}\ee
  Using the equations of motion  (\ref{codiv}), this can be
rewritten as  $ \epsilon_{ab} \partial^\mu   ( \chi_a \pi_\mu^b )
=0 ,$  which is locally solved by writing
 \be
  \epsilon_{\mu\nu}\partial^\nu \theta =
  \epsilon_{ab}\chi_a\pi_\mu^b\;. \label{defchi1}  \ee
We use this equation to define the scalar field  $\theta$, which
we now promote to a degree of freedom to be varied in the action.
(More precisely, as stated earlier, it will play the role of an auxiliary
 variable.)  This
will allow us to recover
 eq. (\ref{defchi1}), as well as
(\ref{do3eom})  from an action principle, through
 variations of $A_\mu$ and $\theta$,    respectively.
  The Lagrangian $\tilde L_0$
is, however, insufficient for this purpose, as we shall need to add a $BF$
or instanton term.
Furthermore, $\alpha$ and $\beta$ now need to be functions of  $\theta$.
[This won't spoil the  result of   (\ref{codiv})  for
the equations of motion.]      The total  Lagrangian is then
\be \tilde L = \tilde L_0 +\epsilon^{\mu\nu}\partial_\mu\theta A_\nu
\label{totdL}\;. \ee  $\theta$ is assumed to be gauge invariant.  Then,
although
$\tilde L$  is then no longer gauge invariant under
(\ref{dgt}), the corresponding action integral is.   Now it easily
follows that variations with respect to $A_\mu$ lead to
(\ref{defchi1}).  Since the latter is associated with an identity,
there are no true dynamical degrees of freedom in $A_\mu$.
 Variations with respect to $\theta$ lead to
\be F= -{{\alpha'}\over 2} (D_\mu \chi)_a  (D^\mu \chi)_a-
 {{\beta'}\over 2}
\epsilon^{\mu\nu}\epsilon_{ab} (D_\mu \chi)_a (D_\nu \chi)_b \;,\ee
the prime indicating a derivative  with respect to $\theta$.
Using  (\ref{ddefpi}), this agrees with (\ref{do3eom}) provided that
\be \alpha'=2\alpha \beta \;,\qquad \beta'=\beta^2-\alpha^2 \;.\ee
    These equations are solved by
\be \alpha=\frac\kappa{\kappa^2+\theta^2}\;,\qquad
 \beta=-\frac{\theta}{\kappa^2+\theta^2} \;,
\label{defab}\ee   $\kappa$ again denoting an arbitrary constant.
From the Hamiltonian analysis (see section 5), we find that this
constant can be identified with the constant $\kappa$ appearing in
 the primary Lagrangian.   It remains to obtain  (\ref{o3eom}).
It follows from another useful identity.
  For this let us invert (\ref{ddefpi}), using (\ref{defab}),
   to solve for $(D_\mu \chi)_a$:
\be  (D_\mu \chi)_a = -\kappa\pi_\mu^a -\theta \epsilon_{\mu\nu}
\epsilon_{ab}
\pi^{\nu b}                          \;.\ee   Now
 take the covariant curl to get
\be \kappa\epsilon^{\mu\nu} (D_\mu \pi_\nu)^a =  \epsilon_{ab}\biggl
(F \chi_b -
  \partial^\mu\theta \pi_\mu^b -\theta (D_\mu \pi^\mu)^b \biggr)  \;.
 \label{covcurpi} \ee
The right hand side is seen to vanish upon imposing the equations of
motion  (\ref{codiv}),  (\ref{do3eom}) and  (\ref{defchi1}), and hence we
recover the equation of motion of the primary formulation (\ref{o3eom}).

We thus have obtained all the dynamical equations of the primary
 formulation.
We note that while the coefficient $\kappa$ of the kinetic energy term
in the standard description is a constant, the analogue in the
dual description goes like $(\kappa +\theta^2/\kappa)^{-1}$.  (A nonlocal
expression for the latter results if we use (\ref{defchi1}) to eliminate
$\theta$.)  This relation between coefficients resembles the
$R\leftrightarrow \frac 1R$
 correspondence in Abelian T-duality.\cite{kysy}
Concerning symmetries, while the $U(1)$ gauge symmetry is obvious
in the dual description, the global $SU(2)$ symmetry
(\ref{glblsym}) is not.  In the dual theory, the latter
 can only be defined
nonlocally.  It should  now be possible to recover all previous
 expressions for the dual Lagrangian by choosing the appropriate
gauge and eliminating the auxiliary variable $\theta$.

\subsection{Higher Order Terms}

It is easy to consider the possibility of adding higher order
terms to the primary Lagrangian (\ref{gstLagr}) or (\ref{orLa}).
For example, we can add the quartic term
\be L_{quar}=\frac\xi{2}F^2 \;,\label{hot}\ee where $\xi$ is
a constant and $F$ is defined in terms of $SU(2)$ group elements
 $g$ through (\ref{psitg}) and
(\ref{instfld}).  Then the equation of motion (\ref{o3eom})
is replaced by
\be \kappa\epsilon^{\mu\nu}(D_\mu \pi_\nu)^a =-\xi \epsilon_{ab}
\partial^\mu F\pi_\mu^b   \;.
\label{o3eompq}\ee    The  identities   (\ref{codiv}) and (\ref{do3eom})
are, of course, unchanged.  One adds precisely
 the same term (\ref{hot})
to the dual Lagrangian  (\ref{totdL}),
 only there $A_\mu$ is treated as an independent
degree of freedom.  Thus   a kinetic energy is now
 associated with
 $A_\mu$.    Its variations leads to
 \be
  \epsilon_{\mu\nu}\partial^\nu (\theta+\xi F) =
  \epsilon_{ab}\chi_a\pi_\mu^b\;, \label{mdefchi1}  \ee  replacing
(\ref{defchi1}).   None of the other  equations of motion for the dual
system are affected.
 Upon substituting (\ref{mdefchi1}),
as well as (\ref{codiv}) and  (\ref{do3eom}), into the identity
 (\ref{covcurpi}),     we once again get  (\ref{o3eompq}).

\section{General Coset Models}
\setcounter{equation}{0}

Here we find the dual Lagrangian description
associated with the primary   Lagrangian  (\ref{gstLagr}),
 for the case of arbitrary
 Lie groups $G$ and
$H\subset G$ .

\subsection{Primary  Formulation}

We begin with the primary Lagrangian (\ref{gstLagr}).  There are now
$M$ equations of motion resulting from variations of $g$, and they
 once again     can  be written as   (\ref{o3eom}).  For this,
 $ \pi_{\mu}^ a  $ is  again defined as in      (\ref{defpi}), while
 \be A^\alpha_\mu=
 (g^{-1}\partial_\mu g)^{M+\alpha}\;.\label{gdefA}      \ee
$A_\mu^\alpha$ transforms as an $H$ connection under (\ref{ggt}).
The covariant derivative is now defined by
\be
(D_\mu\pi_\nu)^a=\partial_\mu\pi^a_{\nu} + c^a _{M+\beta\;\; c}
 A_\mu^\beta  \pi_{\nu}^c\;. \ee
In addition to the equations of motion (\ref{o3eom}),
 we have the generalization of the identities (\ref{codiv}) and
  (\ref{do3eom}) corresponding to $N$ Maurer-Cartan equations:
\beqa
 (D^\mu\pi_\mu)^a & =& \frac12 c^a_{bc}  \epsilon^{\mu\nu}\pi_\mu^b
 \pi_\nu^c\;,\label{gcodiv} \\
F^\alpha & =&{1\over 2}c_{bc}^{M+\alpha}  \epsilon^{\mu\nu}
\pi_\mu^b\pi_\nu^c\;.\label{gdo3eom} \eeqa  Now, in general,
  the covariant
divergence of $\pi_\mu^a$ need not vanish.
$F^\alpha$ is the $H$ curvature,
\be F^\alpha=\epsilon^{\mu\nu}( \partial_\mu A_\nu^\alpha
  +\frac12 \hat{c}
^\alpha_{\beta\gamma}A^\beta_\mu A^\gamma_\nu)\;      \;.\ee

\subsection{Dual Formulation}

For the dual theory, we introduce $N$ scalar fields,
$\chi_a$ and
$\theta_\alpha$, along with the Yang-Mills connection one form
 $A^\alpha$,
which undergo gauge transformations   \beqa
 \delta\chi_a& =&c^b_{M+\alpha\;a}\lambda^\alpha  \chi_b \label{gtochi} \\
\delta \theta_\alpha & =&\hat{c}^\gamma_{\beta \alpha}\lambda^\beta
  \theta_\gamma                                      \label{gtpthe} \\
\delta A^\alpha & =&d\lambda^\alpha+ \hat{c}^\alpha_{\beta \gamma}
A^\beta \lambda^\gamma  \;,\label{gtovp}   \eeqa
    $\lambda =\lambda^\alpha
\hat{e}_\alpha$ being
an infinitesimal element of the Lie algebra ${\cal H}$
 of $H$.
Upon defining the covariant derivative of $\chi_a$ according to
$ (D\chi)_a =d\chi_a +c^b_{a\;M+\alpha}A^\alpha \chi_b \;,$
we can construct the following  Lagrangian density:
\be \tilde L =   -{1\over 2}{\alpha}^{ab} (D_\mu \chi)_a
 (D^\mu \chi)_b-
 {1\over 2} \epsilon^{\mu\nu}\beta^{ab} (D_\mu \chi)_a (D_\nu \chi)_b
 -\theta_\alpha F^\alpha   \;,        \label{gendL} \ee
generalizing (\ref{totdL}).  Now  the dual metric ${\alpha}^{ab}$   and
the antisymmetric tensor $\beta^{ab}$
are functions of $\chi_a$ and      $\theta_\alpha$.
  Neither ${\alpha}^{ab}$   nor ${\beta}^{ab}$ are gauge invariant.
Gauge invariance demands that ${\alpha}^{ab}$
  and $\beta^{ab}$   transform, respectively,
   as symmetric and antisymmetric second rank tensors.
%\beqa \delta  \alpha^{ab} &=&
% ( c^b_{c\;M+\alpha } \alpha^{ac} +  c^a_{c\;M+\alpha} \alpha^{cb} )
% \lambda^\alpha                                       \cr
% \delta  \beta^{ab} &=&
% ( c^b_{c\;M+\alpha } \beta^{ac} +  c^a_{c\;M+\alpha } \beta^{cb} )
% \lambda^\alpha        \;.\eeqa
The form for $\tilde L$, including the $BF$-term, is similar to previous
expressions in the literature.\cite{rtd}  Here we are able to write
down ${\alpha}^{ab}$ and $\beta^{ab}$ without having
to specify coordinates on the original target manifold $G/H$.

To determine the coefficients  ${\alpha}^{ab}$ and $\beta^{ab}$ we
 demand that the equations of motion resulting from variations in the
scalar fields are  (\ref{gcodiv}) and (\ref{gdo3eom}).
 Variations of
$\chi_f$ give the former equation provided that
\beqa \frac{\partial \alpha^{ab}}{\partial \chi_f} &=&
c^{f}_{dc}(\beta^{da}\alpha^{cb}-      \alpha^{da}\beta^{cb})\cr
 \frac{\partial \beta^{ab}}{\partial \chi_f} &=&
c^{f}_{dc}(\beta^{da}\beta^{cb}-      \alpha^{da}\alpha^{cb})\;,
\label{gco}\eeqa
while variations of
$\theta_\alpha$ give the latter equation provided that
\beqa \frac{\partial \alpha^{ab}}{\partial \theta_\alpha} &=&
c^{M+\alpha}_{dc}(\beta^{da}\alpha^{cb}-      \alpha^{da}\beta^{cb})\cr
 \frac{\partial \beta^{ab}}{\partial \theta_\alpha} &=&
c^{M+\alpha}_{dc}(\beta^{da}\beta^{cb}-      \alpha^{da}\alpha^{cb})\;.
\label{gct}\eeqa
$\pi_\mu^a$ is now defined  by  \be  \pi_\mu^a   =
-\alpha^{ab}(D_\mu \chi)_b-\beta^{ab}
\epsilon_{\mu\nu}  (D^\nu \chi)_b \;. \label{gdefpi}
\ee
To write down solutions to   eqs. (\ref{gco}) and (\ref{gct})
we introduce a matrix  $\tilde f$ with matrix elements
which  are linear
functions of  $\theta_\alpha$ and $ \chi_a$,
\be   {\tilde f}(\theta, \chi)_{ij}=
c^c_{ij}\chi_c +c^{M+\alpha}_{ij}\theta_\alpha
 \;.\label{tfsac}\ee
 Then a little work shows that
\be \alpha=\biggl(\kappa\underline{\tt g}-
\frac 1\kappa\underline{\tilde f}
\underline{\tt g}^{-1} \underline
{\tilde f}\biggr)^{-1} \;,   \qquad
  \beta =-\frac 1\kappa \underline{\tt g}^{-1}\underline{\tilde f}
    \alpha \;,\label{tgss}\ee
which generalizes the solution  (\ref{defab}) found for the coefficients
  of the dual nonlinear   $\sigma$-model.
The underline indicates  the   submatrix   obtained by projecting onto
$G/H$.  Thus the matrix $\underline{\tt g}$ consists only
of matrix elements
$\underline{\tt g}_{ab}={\tt g}_{ab},$ $a,b=1,2,...M$.    As before,
    $\kappa$ is
   an arbitrary constant, which we shall identify with the coefficient
$\kappa$ appearing in the primary Lagrangian.  The matrix
 $\alpha$ in (\ref{tgss}) is
 symmetric by inspection, while
antisymmetry for $\beta$ follows after using the identity
\be \underline{\tilde f} \alpha  \underline{\tt g}=
\underline{\tt g}\alpha
\underline{\tilde f}\;.\label{fag=gaf}\ee
  The solution (\ref{tgss}) easily reduces to
(\ref{defab}) when $G=SU(2)$ and $H=U(1)$,   and we thus recover
  the dual formulation of the nonlinear   $\sigma$-model.

 As in the dual formulation of the nonlinear   $\sigma$-model,
 the connections   are nondynamical degree of freedoms and their
variations do not  lead additional conditions.
  Varying $A^\alpha_\mu$ in
 (\ref{gendL})  gives
 \be
  \epsilon^{\mu\nu}(D_\nu \theta)_\alpha =
 c_{b\;M+\alpha}^c\chi_c\pi^{\mu b}\;,\label{codthe} \ee
  generalizing   (\ref{defchi1}).  Here
   $ (D_\mu \theta)_\alpha =   \partial_\mu \theta_\alpha
     + \hat{c}_{\alpha
\beta}^\gamma A^\beta_\mu\theta_\gamma \;. $
  By taking the covariant divergence of both sides of (\ref{codthe})
one gets
\be \hat{c}_{\alpha \beta}^\gamma F^\beta\theta_\gamma =
   c^c_{b \;M+\alpha} (D_\mu \chi)_c \pi^{\mu b} +
  c^c_{b \;M+\alpha}  \chi_c (D_\mu \pi^\mu)^b  \;.\ee   Now
applying the equations of motion (\ref{gcodiv}) and (\ref{gdo3eom}),
this becomes
\be      \frac12 c^i_{ab}      \epsilon^{\mu\nu}\pi_\mu^a  \pi_\nu^b\;
{\tilde f}_{M+\alpha \;i}
=c^c_{b\;M+\alpha} (D_\mu\chi)_c\pi^{\mu b} \;.\label{geid}\ee
Eq. (\ref{geid}) generalizes  (\ref{did}), the latter being
 an identity for the dual
  $O(3)$ nonlinear $\sigma$-model.   It is not too hard to show that
   eq. (\ref{geid})
is also an identity.   The procedure is the same  as before.  One
 inverts (\ref{gdefpi}), solving for $(D_\mu \chi)_a $ using
the solutions (\ref{tgss}) found for the coefficients:
\be  (D_\mu \chi)_a = -\kappa
 {\tt g}_{ab}\pi_\mu^b-\epsilon_{\mu\nu}
  {\tilde f}_{ab}
  \pi^{\nu b}      \label{Dchia}
  \;,\ee     and then substitutes into the right hand side of
  (\ref{geid}).

    Finally, if we  take the covariant curl of
 (\ref{Dchia}) we get another identity
\be \kappa{\tt g}_{ab}\epsilon^{\mu\nu} (D_\mu \pi_\nu)^b =
 c_{M+\alpha \; a}^b F^\alpha \chi_b -
 {\tilde f}_{ab}    (D_\mu \pi^\mu)^b
  -  \biggl(c^c_{ab}(D_\mu\chi)_c +c^{M+\alpha}_{ab}(D_\mu\theta)
 _\alpha\biggr)     \pi^{\mu b}   \;,\ee  which generalizes
 (\ref{covcurpi}).
The right hand side of this equation vanishes upon applying
(\ref{codthe}),
(\ref{gcodiv}) and (\ref{gdo3eom})   and hence we recover  the
equations of motion    (\ref{o3eom})  of the standard formalism.
  We thereby
recover all the dynamical equations of the standard formalism.

\section{Hamiltonian descriptions}
\setcounter{equation}{0}

In this section, we note that the Hamiltonian descriptions
for the primary formulation and dual formulation of $G/H$ models
 are, at least locally, canonically equivalent.

\subsection{Primary Formulation}

The Hamiltonian description corresponding to the primary Lagrangian
is written
on an equivalence class of the loop group $LT^*G$ of the cotangent bundle
$T^*G$.  To describe it we can introduce
a $2N$ dimensional  phase space,  spanned by the fields
$g(x)$ and $K_i(x)$, the latter
 generating, for example,  right translations on $G$, i.e.
\beqa   \{ K_{i} (x),  g (y) \}& =&-g(x)e_i \delta(x-y) \cr
\{ K_{i} (x), K_{j} (y) \}&=&- c_{ij}^k  K_{k} (x)  \delta(x-y)
\;.\label{gltoG}\eeqa  As usual, we assume the Poisson brackets
 between two group elements vanishes.
The physical subspace is $2M$ dimensional, and it
is obtained
after moding out gauge transformations  generated by  the  $N-M$
first class constraints
 \be K_{M+\alpha} \approx 0 \;.\label{gcoTG}\ee
   Up to   Lagrange multiplier
terms involving the constraints (\ref{gcoTG}),  the
 canonical Hamiltonian can be written
\be {\tt H}(g,K^i)=
  \frac12\int dx \biggl(\kappa {\tt g}_{ab}
  (g^{-1}\partial_1 g)^a (g^{-1}\partial_1 g)^b\;+\; {1\over \kappa}
   {\tt g}^{ab}
 K_a K_b    \biggr) \;, \label{gsigH} \ee
where $ {\tt g}^{ab}{\tt g}_{bc}=\delta_c^a \;.  $  The
Hamiltonian (\ref{gsigH}) and constraints (\ref{gcoTG}) define the primary
Hamiltonian description.

\subsection{Dual Formulation}

In the dual description, we start with a $6N-4M$ dimensional
phase space spanned by $\chi_a$, $\theta_\alpha$, $A_\mu^\alpha$ and
their corresponding conjugate momenta.
  $\pi^a=\pi^{0a}$, as defined in (\ref{gdefpi}), are
 the momenta conjugate to $\chi_a$.  If we integrate by parts
  the topological term  in (\ref{gendL}),  we get that   the momenta
 $\pi^{M+\alpha}$  conjugate to $\theta_\alpha$ are constrained by
\be  \pi^{M+\alpha}- A_1^\alpha \approx 0 \;,\label{gdcon1}\ee
 and there are
$2(N-M)$ additional primary      constraints stating
that  the momenta $\Pi^\mu_\alpha$ conjugate to $A^\alpha_\mu$ vanish,
 \be \Pi^\mu_\alpha  \approx 0 \;.\label{gdcon23}\ee
In addition, there are   $ N-M $       secondary
constraints of the form
  \be (D_1\theta)_\alpha +c_{M+\alpha \;b}^c\pi^b
\chi_c\approx 0 \;.\label{gdcon4}\ee     They are analogous to
the Gauss law constraints.
   Up to the constraints, the Hamiltonian is given  by
 \beqa \tilde {\tt H}& =&\frac 1{2}\int dx\;\alpha^{ab}\biggl(
(D_0\chi)_a   (D_0\chi)_b  +   (D_1\chi)_a   (D_1\chi)_b  \biggr) \cr
 & =&\frac 1{2}\int dx\biggl(
  \kappa {\tt g}_{ab}\pi^a \pi^b
   +{1\over \kappa}{\tt g}^{ab}  \;
 [ (D_1 \chi)_a +   {\tilde f}_{ac}  \pi^c ]   \;
 [ (D_1 \chi)_b +   {\tilde f}_{bd}  \pi^d ] \biggr)
 \label{dsigHam}
 \;.\eeqa   To write the second line, we used the identity \be
 \kappa \underline{\tt g} +
\underline{\tilde f}\alpha  \underline{\tilde f}
 =\kappa^2  \underline{\tt g}
\alpha \underline{\tt g} \; ,\label{said}\ee
 which follows from (\ref{tgss}).

To summarize, we get
  $6N-4M$ phase space variables subject to $4(N-M)$
constraints.  The constraints can be evenly divided into a first class
and second class set.       For the
 latter, we take (\ref{gdcon1}) and
  $\Pi^1_\alpha  \approx 0$.  To implement the second class constraints
  we can
go to a $4N-2M$ dimensional reduced phase space spanned by
Dirac variables.
  A choice for
Dirac variables is:   $A_0^\alpha$,
$\Pi^0_\alpha$, $\chi_i$ and  $ \pi^i,\; i=1,2,...,N$, where we define
    \be
 \chi_{M+\alpha}=\theta_\alpha -\Pi^1_\alpha \;.\ee
  The  variables $A_0^\alpha$ and $\Pi^0_\alpha$
are nondynamical and can be eliminated.  This is because
$\Pi^0_\alpha  \approx 0$ are first
class, and
  $A_0^\alpha $ do not appear in
the Hamiltonian.  On the reduced phase space the latter is expressed by
 \be \tilde {\tt H}(\chi_i,\pi_i) =\frac 1{2}\int dx\biggl(
 \kappa {\tt g}_{ab}\pi^a \pi^b
  +{1\over \kappa}{\tt g}^{ab}     \;
 [\partial_1 \chi_a +  c_{ai}^{j}  \pi^i\chi_j ]  \;
 [\partial_1 \chi_b +  c_{bk}^{\ell}  \pi^k\chi_\ell ]
  \biggr)     \;.\label{gdualHam}\ee
  The remaining $2N$ variables
$\chi_i$ and $ \pi^i$ are canonically conjugate, their equal time
Poisson brackets being
 \be\{\chi_i(x),\pi^j(y)\}
=\delta_i^j \delta(x-y) \;.\label{canbra}\ee  They
 are subject to the remaining
 constraints  (\ref{gdcon4}), which can be expressed by
 \be \phi_\alpha = \partial_1\chi_{M+\alpha}
  +c^j_{M+\alpha \; i}\pi^i\chi_j
  \approx 0\;.\label{gfccon}  \ee  They are  first class since
 \be \{ \phi_\alpha (x),\phi_{\beta} (y) \}=-
  \hat{c}_{\alpha\beta}^\gamma \phi_\gamma (x)
   \delta(x-y)          \;.\ee     Thus as in the primary  Hamiltonian
description, the resulting physical subspace is  $2M$ dimensional.  It
is obtained
after moding out gauge transformations,  now generated by (\ref{gfccon}),
from the space spanned by $\chi_i$ and $ \pi^i$.

It is now evident that the two dual Hamiltonian descriptions for coset
models are canonically equivalent, at least locally.
The Hamiltonian  (\ref{gdualHam})  and constraints (\ref{gfccon}) are
obtained
 from  (\ref{gsigH})   and (\ref{gcoTG}) using the canonical
 transformation\cite{prwo} :
\beqa  K_i &\rightarrow&  \partial_1\chi_i +c_{ij}^k\pi^j\chi_k
\label{cantr1}\\
  (g^{-1}\partial_1 g)^i  &\rightarrow&  \pi^i  \label{cantr2} \;.\eeqa
  This is the same
canonical transformation that has  been used to relate the
standard description of the
 $SU(2)$  Principal Chiral Model with its dual description,\cite{loz} and
it is a special case of the canonical transformation linking theories
which are Lie-Poisson dual.\cite{Sfet},\cite{prwo}   We consider the more
general class of canonical transformations in the next section.
Concerning (\ref{cantr1}), using the canonical Poisson brackets
 (\ref{canbra}),
we can easily show that the right hand
side generates the group $G$.
Concerning (\ref{cantr2}), we get that the canonical momenta
$\pi^a$
in the dual theory are identified with $\pi^{0a} $ defined by
  (\ref{defpi})
in the primary formulation.  Similarly, from (\ref{gdcon1})
 $\pi^{M+\alpha}$ gets
identified with $A^\alpha_1$ as defined by (\ref{gdefA})
in the primary formulation.

\subsection{Current Algebra}

The Poisson algebra of currents for the two theories can
 be reexpressed in terms of  an equivalence
 class of Kac-Moody algebras associated with the cotangent bundle
group   $T^*G$.
   The latter is generated by $e_i$,
     along with another $N$ generators
   $e^i$  , which
have commutation relations
 \beqa     [e^i,e_j]&=& c^i_{jk} e^k  \cr    [e^i,e^j]&=&0 \;.
 \label{genTsG}\eeqa
We can introduce a nondegenerate
  invariant scalar product on the Lie algebra spanned
$e^i$ and $ e_i $:
 $$< e^i|e_j>=\delta_j^i\;,\qquad < e^i| e^j>=   <e_i|e_j>=0 \;.$$
In terms  of the  phase space variables $K_i$ and $g$ of the primary
Hamiltonian description, we now define the current
\be v =  K_ie^i  +       g^{-1}\partial_1 g        \;.\label{vitgK}\ee
Its resulting Poisson algebra is
the central extension of the loop  group of $T^*G$:
\be   \{\vonex,\vtwoy\} =-[C,\vonex]\delta(x-y) + C\partial_x \delta(x-y)
\;,\label{lgoD}\ee
where we  use tensor product notation,
the $1$ and $2$ labels referring to two separate vector spaces, with
$\vonex=v(x)\otimes \BI$,  $\vtwoy=\BI\otimes v(y)$, and
$\BI$ being the unit operator.
$C$ in (\ref{lgoD}) is a constant adjoint invariant tensor defined by
 $C=e^i \otimes e_i + e_i \otimes e^i \;.$   Of course, the same algebra
results if we express $v$ in terms of the dual phase space variables
$\chi_i$ and $\pi^i$ according to
\be v= (\partial_1\chi_i +c_{ij}^k\pi^j\chi_k  ) e^i + \pi^ie_i \;.\ee
    The equivalence class
results after moding out the $H$ gauge
 transformations generated by (\ref{gcoTG}),
i.e.     \be <\hat{e}_{\alpha}|v(x)> \approx 0 \;.\label{vsubal}\ee
%Infinitesimal guage transformation are of the form
%\be \delta \vonex =    <[C,\lambdaonex]|\vtwox>_2 -\partial_x \lambdaonex
%\;,\ee where $\lambda(x)= \lambda^\alpha(x) \hat{e}_\alpha$ and
% $<\;|\;>_2$ indicates that the scalar product is on
%  the second vector space in the tensor product.
In terms of the currents, the Hamiltonian   is
 \be  {\tt H}(v) =\frac 1{2}\int dx  <
 \kappa {\tt g}_{ab}  e^a \otimes e^b +
 {1\over \kappa}{\tt g}^{ab}   e_a \otimes e_b\;
 |\;v(x)\otimes v(x)>
     \;,\label{Hamv}\ee  and it
is preserved under such transformations.

\section{Lie-Poisson Generalizations}
\setcounter{equation}{0}

The kind of duality we have been studying so far can be regarded
as traditional nonabelian T-duality.  It has also been referred
to as semiabelian duality.  This is because while the fields
of the primary theory take values in a nonabelian group,
the fields in the dual theory are associated with an $N$ dimensional
 Abelian group, which we shall denote by $\tilde G$.
   It is the group generated by $e^i$, $i=1,2,...N$.
One may then  ask whether or not it is possible to generalize
our formalism
to the case of nonabelian  $\tilde   G$.  The answer is yes, and
we claim that there is a natural generalization of the Lagrangian
 (\ref{gendL}) to one written on $T\tilde G$, where the function
 $\tilde f$ appearing in the dual metric and antisymmetric tensor becomes a
  nonlinear   function of the fields.
  \footnote{For another approach  to this subject see
 refs. \cite{klm96} and \cite{sf2}.}

\subsection{Generalizing the Current Algebra}

 The natural choice
for  $\tilde G$ is that it  be dual of $G$ in the Lie-Poisson sense.
This means that, $G$ and $\tilde G$  are maximally isotropic subgroups
of a group $ G_{D}$ known as the Drinfeld double.\cite{drin}
Then $dim[G] =dim[\tilde G]$  $ =\frac12 dim[G_D] $.
$G_D$ is generated
by all of $e_i$ and $e^i$.  In
 the case where $e^i$ and $e_i$ satisfy commutation
relations    (\ref{crG})   and  (\ref{genTsG}), $G_ D$ is
 $T^*G$.   More generally, if we denote by $c^{ij}_k$
  the structure constants of $\tilde G$,  we can have
\beqa     [e^i,e_j]&=& c^i_{jk} e^k - c^{ik}_je_k  \cr
  [e^i,e^j]&=&c^{ij}_k e^k \;, \eeqa
  along with (\ref{crG}).
Now in addition   to the usual Jacobi identities for
the Lie algebras associated with $G$
and $\tilde G$, we have
\be c^k_{ij}c^{\ell m}_k = c^\ell_{ki} c^{mk}_j  -c^m_{ki} c^{\ell k}_j
-  c^\ell_{kj} c^{mk}_i + c^m_{kj} c^{\ell k}_i  \;.\ee

 We  next ask, given the above generalization of the Lie algebra
 ${\cal G}_D$ spanned by $e^i$ and $e_i$,  whether
 or not the Hamiltonian dynamics
 of the previous section
can still be utilized to  describe $G/H$
 coset models.   More specifically, using the current algebra
  (\ref{lgoD}), which now  is the Kac-Moody algebra
associated with $G_D$, is the Hamiltonian (\ref{Hamv}) still
(at least, weakly)
invariant under gauge transformations generated by (\ref{vsubal})?
It is easy to check that this is indeed the case provided that we have
the following restriction on the structure constants of $\tilde G$:
    \be       c^{bc}_{M+\alpha}    = 0  \;,\label{rdc}
    \ee  along with the conditions
(\ref{stco}) and (\ref{metr}).
This then means that, in addition to $G$ having an $N-M$ dimensional
subgroup $H$,
 $\tilde G$ must have an $M$ dimensional subgroup, which we denote by
  $\tilde H$.  The latter is generated by $e^a$, $a=1,2,... M$.

An example  of a Drinfeld group $G_D$ with this kind of structure for
the maximally isotropic subgroups is the Lorentz group or $SL(2,C)$.
There we can take $G=SU(2)$ and $\tilde G=SB(2,C)$, the Borel group.
The structure constants for the former are $c^k_{ij}=\epsilon_{ijk}$,
and for the latter are  $c_k^{ij}=\epsilon_{ij\ell}\epsilon_{\ell k3}$.
$e_3$ generates a  $H=U(1)$ subgroup of $SU(2)$, while $e^1$ and $e^2$
generate a two dimensional Abelian subgroup $\tilde H$ of   $SB(2,C)$.
Since $SU(2)/U(1)$ is the target space for
 the $O(3)$ nonlinear  $\sigma$-model,
we can thus expect to find another dual description of
 this model,
 this time in terms of fields spanning $SB(2,C)$.    As we shall see
below, the primary Lagrangian then corresponds to (\ref{plLagr}).

More generally,  instead of restricting to the dynamics resulting from
the Hamiltonian  (\ref{Hamv}), we could
 consider an arbitrary
quadratic Hamiltonian as is done in \cite{prwo}, i.e.
 \be  {\tt H}(v) =\frac 1{2}\int dx  <
 {\Phi}_{ij}  e^i \otimes e^j +
 {\Gamma}^{ij}   e_i \otimes e_j  + 2{\Theta}_i^{\;\;j}   e^i \otimes e_j
 |\;v(x)\otimes v(x)>
     \;,\label{gqHamv}\ee where
${\Phi}_{ij}$, ${\Gamma}^{ij}$ and ${\Theta}_i^{\;\;j}$ are constant
matrices (the first two being symmetric).  We must impose
 a number of conditions  on the constant matrices
 in order to have a local $H$ invariance.  They are:
$$   {\Phi}_{M+\alpha \; i}  = {\Theta}_{M+\alpha}^{\quad b} =
 {\Phi}_{ bj} c^b_{k \; M+\alpha} + {\Phi}_{ bk} c^b_{j \; M+\alpha}
 =0
$$  \be      {\Phi}_{ cj} c^{bc}_{ M+\alpha}- {\Theta}_c^{\;\;b}
c^c_{j\; M+\alpha} +     {\Theta}_j^{\;\;c}c^b_{c\; M+\alpha}   =0 \ee
$$ {\Gamma}^{ib} c_{i \; M+\alpha}^c
  +{\Theta}_a^{\;\;b}c^{ca}_{M+\alpha} +
   {\Gamma}^{ic} c_{i \; M+\alpha}^b
     +{\Theta}_a^{\;\;c}c^{ba}_{M+\alpha}
 =0 \;.$$   We will not examine this more general possibility below,
but rather restrict to Hamiltonian dynamics given by (\ref{Hamv}).

\subsection{Primary Formulation}

Even though $G_D$ is no longer restricted to being $T^*G$,
the current algebra
 (\ref{lgoD})  can still be realized in terms of fields valued in
$T^*G$, i.e. $g(x)$ and $K_i(x)$, satisfying (\ref{gltoG}).\cite{prwo},
\cite{Sfet}  Now,
 however, for nonzero $c^{ij}_k$
  the expression (\ref{vitgK}) for the currents
$v$ in terms of   $g(x)$ and $K_i(x)$ must be modified\footnote
{Alternatively, we can write $v(x)$ in terms of left generators on $G$,
as is done in \cite{prwo}.}:
\be v =  K_ie^i  +f(g)^{ij} K_j e_i +  g^{-1}\partial_1 g
      \;,\label{rvvitgK}\ee     where
 the matrix valued function $f(g)$ is defined by
\be f(g)^{ij}=<ge^i g^{-1}\otimes ge^j g^{-1}|e^k\otimes e_k> \;.\ee
$f(g)$ is antisymmetric.   This follows from
\beqa  f(g)^{ij} + f(g)^{ji}=<ge^i g^{-1}\otimes ge^j g^{-1}|C>
&=&<  e^i\otimes e^j| \gone^{-1}\gtwo^{-1} C\gone \gtwo >    \cr
&=&<  e^i\otimes e^j|  C >=0  \;.\eeqa
The semiabelian  case discussed in the previous section
 corresponds to setting $f=0$.
   To verify  that  the Poisson algebra
 (\ref{lgoD})  applies even when $f\ne 0$, we can first
compute the Poisson bracket of  $K_i$ with $f(g)^{jk}$:
\be \{K_i(x), f(g)^{jk}(y)\} =\biggl( -c^{jk}_{i} +
c^j_{i\ell} f(g)^{\ell  k}(y)  -c^k_{i\ell} f(g)^{\ell j}(y)\biggr)
\delta(x-y) \;. \ee  Eq. (\ref{lgoD}) then follows after using the
identity
\be c^{[ij}_\ell f^{k]\ell} =  f^{\ell [i} c^j_{\ell m} f^{k]m}  \;,
\label{idfcaf}\ee
the brackets indicating antisymmetrized indices.
This  identity can be proved from the result that the
scalar product of  the commutator
of two $\tilde G$ generators, which can be expressed according to
$$e ^i = f(g)^{ij} e_j + <ge^i g^{-1}| e_j> g^{-1}e^j g \;,  $$ with
a third $\tilde G$ generator is zero.

Writing  the Hamiltonian (\ref{Hamv}) in terms of $g$ and $K_i$,
 we now get
\be {\tt H}(g,K^i)=
  \frac12\int dx \biggl(\kappa {\tt g}_{ab}  \;
  [(g^{-1}\partial_1 g)^a +f(g)^{ac} K_c] \;
  [(g^{-1}\partial_1 g)^b +f(g)^{bd} K_d]\; +\; {1\over \kappa}
    {\tt g}^{ab}
 K_a K_b    \biggr) \;,  \ee generalizing (\ref{gsigH}).
 The constraints
are still given by  (\ref{gcoTG}).  The corresponding Lagrangian is
 (\ref{plLagr}),
        where  the metric $\gamma$ and antisymmetric tensor $\rho$
  are
\be \kappa\gamma=\biggl(\frac 1\kappa \underline{\tt g}^{-1} -\kappa
\underline f
\underline{\tt g}\underline f \biggr)^{-1} \;,   \qquad
  \rho =-\kappa \underline{\tt g}\underline  f\gamma
     \;.\label{gamrho}\ee   As before
the underline indicates that we are restricting to the submatrix
with elements in $G/H$.
      Since, provided (\ref{rdc}) holds,
the Hamiltonian dynamics is  invariant under $H$ gauge transformations,
so is the Lagrangian dynamics.   It can be checked that the solution
(\ref{gamrho})
for the metric $\gamma$ and the antisymmetric tensor $\rho$ is consistent
with the
gauge transformation property  (\ref{togam}).   Thus, by construction,
  (\ref{plLagr}) is invariant under  (\ref{ggt}).  On the other hand,
 it is no longer
invariant under global $G$ transformations (\ref{glblsym}) (except,
 of course, when $f=0$).
In the semiabelian case, i.e. $f=0$, we
easily recover the Lagrangian
(\ref{gstLagr}) from (\ref{plLagr}), since
 when $f$ is zero, so is $\rho$, while $\gamma$ reduces to the constant
  metric     $\underline{\tt g}$.

\subsection{Dual Formulation}

Concerning  the  dual
Hamiltonian formulation,  we introduce phase space variables
 $\tilde g(x)\in \tilde G$ and $\tilde K^i(x)$   with
values in $T^*\tilde G$.  The nonzero Poisson brackets are
\beqa   \{ \tilde K^{i} (x),  \tilde g (y) \}& =&-\tilde g(x)e^i
 \delta(x-y) \cr
\{ \tilde K^{i} (x), \tilde K^{j} (y) \}&=&- c^{ij}_k  \tilde K^{k} (x)
 \delta(x-y)
\;.\label{gltoH}\eeqa
Just as the current algebra
 (\ref{lgoD})  can  be realized
  in terms of fields valued in $T^*G$, so  can it be realized
  in terms of fields valued in $T^*\tilde G$.  In analogy to
  (\ref{rvvitgK}), we have
  \be v =  \tilde K^ie_i +\tilde f(\tilde g)_{ij}\tilde K^j e^i   +
    \tilde g^{-1}\partial_1 \tilde g
     \;,\label{vitdgK}\ee  where we replace the previous
      definition of $\tilde f$
given in (\ref{tfsac}) by
\be \tilde f(\tilde g)_{ij}=<\tilde g e_i \tilde g^{-1}\otimes
\tilde  ge_j\tilde  g^{-1}|e_k\otimes e^k> \;.\label{nadft}\ee
To recover the previous definition, we can write
 $\tilde g=\exp(\chi_i e^i)$ and set $c^{ij}_k=0$.  Then
  $ \tilde f(\tilde g)_{ij}  =          c_{ij}^k \chi_k \;.$
Returning to  the  general case, upon
 comparing with (\ref{rvvitgK}), we see that the duality transformation
is now given by
\beqa  K_i &\rightarrow& (\tilde g^{-1} \partial_1\tilde g)_i +
\tilde f(\tilde g)_{ij} \tilde K^j     \\
  (g^{-1}\partial_1 g)^i +f^{ij}(g) K_j &\rightarrow&  \tilde K_i
   \;,\eeqa   generalizing the canonical transformations (\ref{cantr1})
and  (\ref{cantr2}) of the semiabelian case.
As is also true in the semiabelian case, the canonical transformation
between phase space coordinates, here $(g, K_i)$ and
$(\tilde g,\tilde K^i)$, can only be defined locally.

Writing  the Hamiltonian (\ref{Hamv}) in terms of the $T^*\tilde G$
  variables,
  we now get
 \be \tilde {\tt H}(\tilde g,\tilde K_i)=
   \frac12\int dx \biggl(\kappa {\tt g}_{ab}  \tilde K^a
    \tilde K^b  \; +\; {1\over \kappa}   {\tt g}^{ab}  \;
   [(\tilde g^{-1}\partial_1 \tilde g)_a +\tilde f(\tilde g)_{ai}\tilde
    K^i] \;
   [(\tilde g^{-1}\partial_1 \tilde g)_b +\tilde f(\tilde g)_{bj}\tilde
    K^j]            \biggr) \;,\label{lpdh}  \ee
     while the constraints are
 \be \phi_\alpha= (\tilde g^{-1}\partial_1 \tilde g)_{M+\alpha}
  +\tilde f(\tilde g)_{M+\alpha \;i}\tilde       K^i \approx 0 \;.
  \label{lpdc}\ee   From (\ref{lpdh}) and  (\ref{lpdc})
we easily recover the Hamiltonian (\ref{gdualHam})
  and constraints  (\ref{gfccon}) for the dual system in
the semiabelian case.

As gauge transformations are
generated by   the constraints (\ref{lpdc}),
 we can now see how the gauge group    $H$ acts on the dual group
  $\tilde       G$, and its corresponding Lie algebra $\tilde {\cal G}$.
  (Note that, in general, $H$ is not a subgroup of
$\tilde G$.)
    These transformations  are   $ nonlinear$.
Infinitesimal  variations     $\delta   \tilde g  $ of
$\tilde g\in\tilde G$  are of the form:
\be \delta \tilde g=  \lambda^\alpha \tilde f(\tilde g)_{M+\alpha\; i}\;
\tilde g e^i = \tilde g \lambda \; -\;
<\lambda |\tilde g^{-1} e^i \tilde g>
e_i \tilde g
  \;,\label{gvotg}\ee
where $\lambda=\lambda ^\alpha \hat{e}_\alpha$ are infinitesimal elements
of the Lie algebra ${\cal H}$
 of $H$.      (\ref{gvotg})   gives
the Lie-Poisson generalization of (\ref{gtochi}) and (\ref{gtpthe}).
For the purpose of writing down the corresponding dual Lagrangian,
we introduce the `covariant derivative'  $D$ on $\tilde G$.
For this we first note that as a result of (\ref{gvotg}), left invariant
one forms  on $\tilde G$
 undergo the following gauge variations:
\be \delta(\tilde g ^{-1} d \tilde g)_i = d\lambda^\alpha \tilde
f(\tilde g)_{M+\alpha\; i} +\lambda^\alpha
 \biggl(c^j_{M+\alpha \; i} - c^{jk}_{M+\alpha}
\tilde f(\tilde g)_{ki}\biggr)
(\tilde g ^{-1} d \tilde g)_j \;.\label{gtoof}\ee
Let us now define the covariant derivative for left invariant one forms
 according to:
\be (\tilde g ^{-1} D \tilde g)_i =(\tilde g ^{-1} d \tilde g)_i
+  \tilde f(\tilde g)_{i\;M+\alpha} A^\alpha \;,\label{nlcd}\ee
  $ A^\alpha $
being an $H$ connection one form.     Then using  (\ref{gtovp}),
(\ref{gvotg}) and (\ref{gtoof}),  it undergoes the gauge variations
 \be \delta(\tilde g ^{-1} D \tilde g)_i =
\lambda^\alpha
  \biggl(c^j_{M+\alpha \; i} - c^{jk}_{M+\alpha}
 \tilde f(\tilde g)_{ki}\biggr)
 (\tilde g ^{-1} D \tilde g)_j \;.\label{gtocof}\ee   For this we also
  needed   the `dual' of the identity  (\ref{idfcaf}), i.e.
\be c_{[ij}^\ell \tilde f_{k]\ell} =  \tilde f_{\ell [i} c_j^{\ell m}
\tilde f_{k]m}  \;. \ee   We can use (\ref{gtocof}) to define
`covariant' vectors on  $\tilde {\cal G}$.  In other
 words, a `covariant' vector $w=w_ie^i$ having values
 in  the Lie algebra of $\tilde G$ undergoes $H$ variations of the form
 \be \delta w = [w,\lambda]\; - \;
<[w,\lambda] |\tilde g^{-1} e^i \tilde g>
\tilde g^{-1} e_i \tilde g  \label{covaw} \;.\ee
          The second term  introduces a nonlinear contribution to the
 transformation.  It vanishes in the
semiabelian case, and there we recover the usual adjoint action.
   (\ref{covaw})    agrees with (\ref{gtocof})
for $w=\tilde g ^{-1} D \tilde g$.   Moreover, if $x=x^\alpha
 \hat{e}_\alpha\in {\cal H} $ gauge
 transforms under the adjoint action of $H$, i.e.
$\delta x = [x,\lambda]$,  then    $
w=\tilde f(\tilde g)_{i\;M+\alpha} x^\alpha e^i$  is a covariant vector
in  $\tilde {\cal G}$.

We now claim that the dual Lagrangian is
\beqa
 \tilde L & =&   -{1\over 2}{\alpha}^{ab} (\tilde g)\;
(\tilde g ^{-1} D_\mu \tilde g)_a
 (\tilde g ^{-1} D^\mu \tilde g)_b\;-\;
 {1\over 2} \epsilon^{\mu\nu}\beta^{ab}(\tilde g)
  (\tilde g ^{-1} D_\mu \tilde g)_a (\tilde g ^{-1} D_\nu \tilde g)_b \cr
 & &\;+ \; \epsilon^{\mu\nu} (\tilde g ^{-1} \partial_\mu \tilde g)_
 {M+\alpha} A^\alpha_\nu \;-\; \frac12
 \epsilon^{\mu\nu} \tilde f(\tilde g)_{M+\alpha \;
 M+\beta}  A^\alpha_\mu   A^\beta_\nu
    \;.    \label{lpgendL} \eeqa
This is a straightforward generalization of (\ref{gendL}).
The first line gives the coupling to the target metric  and
antisymmetric tensor, while        the second line  generalizes
of the topological term $ -\theta_\alpha F^\alpha  $ appearing
in  (\ref{gendL}).   Now, in general, the latter is not separately gauge
invariant (not even up to
 a total divergence).  Rather, the gauge symmetry is hidden, and
 follows only after considering all terms in the action.

 To check that the expression (\ref{lpgendL})
  is correct, and  also to determine
the functional form for the dual metric $\alpha^{ab}(\tilde g)$ and
antisymmetric tensor $\beta^{ab}(\tilde g)$,
we apply the  Dirac  procedure
 to recover the above Hamiltonian description.
The analysis is essentially the same as in the semiabelian case.
 Following \cite{book},
  let us parametrize the $\tilde G$ group elements by  $N$
variables, which we denote by $\xi^i$; $\tilde g= \tilde g(\xi^i)$.
The momenta $\pi_i$
 conjugate to these variables  are then
\be \pi^i= (\tilde g^{-1} \frac{\partial\tilde g}{\partial \xi_i})_j\;
\biggl(\delta^j_a\;\alpha^{ab}(\tilde g)(\tilde g^{-1}D_0 \tilde g)_b\;-\;
\delta^j_a\;\beta^{ab}(\tilde g)(\tilde g^{-1}D_1 \tilde g)_b \;+\;
\delta^j_{M+\alpha} A^\alpha_1 \biggr)\;.\label{dcm}\ee
   Right generators
$\tilde K^i$ are obtained after introducing matrices
 $N^i_{\;\;j}(\tilde g)$,
defined by \be     e^i=  N^i_{\;\;j} (\tilde g) \;
\tilde g^{-1} \frac{\partial\tilde g}{\partial \xi_j} \;,\ee and writing
\be \tilde   K^i = N^i_{\;\;j} (\tilde g)  \pi ^j \;.\ee  As a result,
we get that $\tilde g$ and $\tilde K^i$ satisfy Poisson brackets
(\ref{gltoH}).  From (\ref{dcm}),
 \be  \tilde   K^a
=\alpha^{ab}(\tilde g)(\tilde g^{-1}D_0 \tilde g)_b\;-\;
\beta^{ab}(\tilde g)(\tilde g^{-1}D_1 \tilde g)_b     \;,\label{tiKa} \ee and in
analogy to (\ref{gdcon1}), we get the primary constraints
\be  \tilde K^{M+\alpha}- A_1^\alpha \approx 0 \;.\label{lpcon1}\ee
Also, as in the semiabelian case, there are the additional primary
      constraints (\ref{gdcon23}) stating
that  the momenta $\Pi^\mu_\alpha$ conjugate to $A^\alpha_\mu$ vanish.
 The    secondary, or Gauss law,
 constraints (\ref{gdcon4}) now become
   \be (\tilde g D_1\tilde g)_{M+\alpha} + \tilde f(\tilde g)
   _{M+\alpha \;b}\tilde K^b\approx 0 \;,\label{lpcon4}\ee
   while the Hamiltonian is given  by
  \be \tilde {\tt H} =\frac 1{2}\int dx\;\alpha^{ab}(\tilde g)\;\biggl(
 (\tilde g^{-1} D_0\tilde g)_a   (\tilde g^{-1} D_0\tilde g)_b  +
  (\tilde g^{-1} D_1\tilde g)_a   (\tilde g^{-1} D_1\tilde g)_b  \biggr)
  \label{HigiDg} \ee    We can  use (\ref{tiKa}) to  reexpress
   $(\tilde g^{-1} D_0\tilde g)_a$
in $\tilde {\tt H}$  in terms of phase space coordinates.

Now let us proceed with the constraint analysis.  As in the
semiabelian case, $A_0^\alpha$  are  nondynamical, and thus  they
may be eliminated  and
 $\Pi^0_\alpha $ may be strongly set equal to zero.  The constraints
 (\ref{lpcon1}), along
  with $\Pi^1_\alpha \approx 0$, form a second class set, and they
are eliminated by going to the constrained surface.  For this we
 introduce Dirac variables  $ K^{*i}$ and $    g^*$,
\beqa K^{*i} &=& \tilde K^i - c^{\alpha i}_j   \tilde K^j \Pi^1_\alpha
\cr  g^* &=& \tilde g \; \varrho (\Pi^1 )\;,\qquad
\frac{\partial \varrho}{\partial \Pi^1_\alpha }\varrho ^{-1}
  = - e^\alpha \;,  \eeqa
where   $\varrho (\Pi^1 )$ takes values in $\tilde G$ and we assume it
to be weakly equal to the identity element.  It is easily
checked that these variables have (weakly) zero Poisson brackets with
(\ref{lpcon1})  and  $\Pi^1_\alpha \approx 0$.   Moreover, the
algebra of Dirac variables is identical to that of
$ \tilde K^i $ and $ \tilde g$.  Thus  the  cotangent bundle algebra
is preserved upon projecting onto the constraint surface, and we
recover  (\ref{gltoH}).  Finally, we note that the remaining
set of constraints (\ref{lpcon4}) agree with (\ref{lpdc}) upon using
(\ref{lpcon1}).  The Hamiltonian (\ref{HigiDg}) is equivalent to
(\ref{lpdh}) provided that the dual metric $\alpha$ and antisymmetric
 tensor $\beta$ are once again given by the expressions (\ref{tgss}).
 Now, of course, $\tilde f$ is, in general, not  a linear function of the
scalar fields, but rather is given by (\ref{nadft}).  We note that
the identity (\ref{fag=gaf}) still holds in this case.

To summarize,  the solution for the metric $\kappa \gamma$ and
 antisymmetric tensor
$\kappa\rho$   (\ref{gamrho}) in the primary Lagrangian
 are expressed in terms of the functions $f(g)$, while their dual
 counterparts $\alpha$ and $\beta$ (\ref{tgss}) expressed in terms of
 $\tilde f(\tilde g)$.  By comparing these expressions, we see that
a duality transformation corresponds to
\be \kappa {\tt g} \leftrightarrow \frac 1\kappa {\tt g}^{-1}\;,\qquad
 f(g)    \leftrightarrow  \tilde f(\tilde g) \;.\ee
The duality transformation for coset theories is, however, more involved
then this, since it also requires  replacing the ordinary
derivatives appearing in the primary Lagrangian with covariant derivatives
(\ref{nlcd}) in the
dual Lagrangian, as well as, including
 a topological term.
The Lagrangian  (\ref{plLagr})  and its dual  (\ref{lpgendL})
are valid for any pair $(G,\tilde G)$ of isotropic subgroups of the
Drinfeld double group $G_D$, and for any subgroup $H$ of $G$, provided
the   restrictions  (\ref{metr}), (\ref{stco}), (\ref{rdc})
on the structure constants and invariant metric on $G$ hold.
Conditions (\ref{metr}) and (\ref{stco}) were necessary already in the
 semiabelian case.
Condition (\ref{rdc}) implied that $\tilde G$ has a subgroup $\tilde H$
of dimension $dim[\tilde H]=dim[G] -dim[H]$.  It is curious that we
end up with complimentary subgroups in $G$ and $\tilde G$.
  In our formalism,   however,
 $\tilde H$ plays a different role than $H$, since it is not gauged.
Of course, provided that there is a metric on $\tilde G$ which
 is invertible and block
diagonal, we can reverse the roles of the two subgroups, gauging
 $\tilde H$ and not $H$.  Then the gauge group
   would have a linear action
on $\tilde G$, and a nonlinear one on $G$.  A more intriguing possibility
would be if we gauged part (i.e. a subgroup) of $H$  and part of
$\tilde H$.  Then the primary and dual Lagrangians would have both a
linear and nonlinear realization of the full gauge symmetry.

 \end{document}